Title: Using Statistical Precision Medicine to Identify Optimal Treatments in a Heart Failure Setting

Authors: Arti Virkud, Jessie K. Edwards, Michele Jonsson Funk, Patricia Chang, Abhijit V. Kshirsagar, Emily W. Gower*, Michael R. Kosorok*
*co-senior authors

**Abstract**
Identifying optimal medical treatments to improve survival has long been a critical goal of pharmacoepidemiology. Traditionally, we use an average treatment effect measure to compare outcomes between treatment plans. However, new methods leveraging advantages of machine learning combined with the foundational tenets of causal inference are offering an alternative to the average treatment effect. Here, we use three unique, precision medicine algorithms (random forests, residual weighted learning, efficient augmentation relaxed learning) to identify optimal treatment rules where patients receive the optimal treatment as indicated by their clinical history. First, we present a simple hypothetical example and a real-world application among heart failure patients using Medicare claims data. We next demonstrate how the optimal treatment rule improves the absolute risk in a hypothetical, three-modifier setting. Finally, we identify an optimal treatment rule that optimizes the time to outcome in a real-world heart failure setting. In both examples, we compare the average time to death under the optimized, tailored treatment rule with the average time to death under a universal treatment rule to show the benefit of precision medicine methods. The improvement under the optimal treatment rule in the real-world setting is greatest (additional ~9 days under the tailored rule) for survival time free of heart failure readmission.

**Introduction**
Statistical precision medicine has attracted the attention of clinical audiences for the past few decades.[1–3] While the phrase "precision medicine" currently has multiple meanings, we define it as the generation of individualized treatment decision rules when not all patients experience the same impact from treatment.[4] Precision medicine relies on a combination of mathematical and engineering fields using public health and clinical data. Some precision medicine methods have leveraged reinforcement learning to identify optimal decision rules for individual patients.[5–12] These methods exist as a bridge between statistical inference and artificial intelligence. While these methods have existed for over a decade, they are relatively new to epidemiology and might provide benefit for evaluating effect measure modification.

Loop diuretics are a critical therapy for over 6 million American adults that have heart failure. There are three loop diuretics that are commonly prescribed: furosemide, torsemide, and bumetanide. Currently, there is conflicting evidence regarding the optimal loop diuretic choice.[9,13–15] Bioavailability, the amount of drug that enters the circulatory system after being administered, differs across these three loop diuretics, and this difference may partially explain the effect heterogeneity of these treatments.[15–17] We do not know which patient characteristics contribute to this heterogeneity. The presence of known effect heterogeneity (different individuals benefit from treatment) with unknown causes (the differences between individuals that do benefit from those who do not) makes loop-diuretic prescribing an ideal setting to explore how a precision medicine approach may improve outcomes.

The objectives of this study were to demonstrate the benefits of a tailored rule versus a one-treatment-fits-all rule and to find the optimal loop diuretic (furosemide or torsemide) treatment rule at hospitalization discharge. We used three robustly-performing precision medicine methods on Medicare claims data to identify an optimal loop diuretic rule: efficient augmentation and relaxation learning[18], random forest[19], and residual weighted learning[6]. These three methods (learners) are verified approaches that are also computationally efficient to calculate. In this study, the learners will generate an optimal treatment rule as a treatment recommendation of either furosemide or torsemide for each individual.

**Toy Example**
To demonstrate the benefit of a tailored rule, we first examine a hypothetical example where effect heterogeneity leads to differences in risk reduction based on choice of intervention. Imagine a population of hospitalized heart failure patients who are prescribed either furosemide or torsemide at discharge. We use their ejection fraction to classify them into two groups: reduced vs. preserved. In this hypothetical example, individuals with reduced ejection fraction respond better to torsemide than individuals with preserved ejection fraction.
We tabulated this hypothetical population of heart failure patients by history of ejection fraction status (preserved vs. reduced), medication prescribed (furosemide vs. torsemide), and 1-year survival (Table 1). Interpretations from an average treatment effect estimation will yield a one-treatment-fits-all rule. Here, we calculate the absolute risk of all-cause mortality in the population under three rules: everyone received furosemide, everyone received torsemide, and a tailored intervention optimized by ejection fraction history (Table 2).

**Table 1: 1-year risk of mortality in a heart failure hospitalized population treated with furosemide or torsemide at discharge, stratified by ejection fraction history**

| Loop Diuretic | Reduced Ejection Fraction | | | | Preserved Ejection Fraction | | | | Total | | | |
|---|---|---|---|---|---|---|---|---|---|---|---|---|
| | Died | Alive | Total | Mortality Risk | Died | Alive | Total | Mortality Risk | Died | Alive | Total | Overall risk |
| **Furosemide** | 7000 | 2000 | 9000 | 0.78 | 2000 | 6000 | 8000 | 0.25 | 9000 | 8000 | 17000 | 0.53 |
| **Torsemide** | 100 | 1000 | 1100 | 0.09 | 1500 | 250 | 1750 | 0.86 | 1600 | 1250 | 2850 | 0.56 |
| **Total** | 7100 | 3000 | 10100 | 0.70 | 3500 | 6250 | 9750 | 0.36 | 10600 | 9250 | 19850 | 0.53 |

**Table 2: Absolute risk under each rule in 1-modifier setting**

| Rule | Absolute 1-year risk of mortality |
|---|---|
| Everyone receives Furosemide | 0.52 |
| Everyone receives Torsemide | 0.47 |
| Tailored rule | 0.17 |

If we developed an intervention based on the average treatment effect, we would recommend that everyone receive furosemide, since the mortality risk is three percentage points higher among those receiving torsemide compared to those receiving furosemide (Table 1, overall risk). However, by stratifying the tables by ejection fraction, we see that individuals with reduced ejection fraction respond better to torsemide, and the absolute risk is highest if the entire population is prescribed furosemide. A tailored rule reduces the risk of death from 0.53 to 0.17 in the entire hypothetical population (Table 2), highlighting the benefit of a tailored rule in the presence of effect heterogeneity due to ejection fraction history.

While the example above highlights the benefit of a tailored rule, this result can be achieved easily by stratifying the average treatment effect. In practice, we often see results stratified by a single, known effect-measure modifier. However, in the real world, multiple patient characteristics may influence the relationship between treatment and outcome. In a second hypothetical example, we examine the benefits of a tailored rule with three known effect modifiers. The size of the total population and the reduced and preserved ejection fraction subgroups are the same as in the above example, but we further stratified by two additional modifiers: pre-hospitalization history of type 2 diabetes and chronic kidney disease.

**Table 3: 1-year risk of mortality in a heart failure hospitalized population treated with furosemide or torsemide at discharge, stratified by ejection fraction, type 2 diabetes (T2DM), and chronic kidney disease (CKD) history and absolute risks of intervention rules**

|  |  |  | Reduced Ejection Fraction | | | Preserved Ejection Fraction | | |
|---|---|---|---|---|---|---|---|---|
|  |  |  | Died | Alive | Total | Died | Alive | Total |
| Has T2DM | Has CKD | Furosemide | 3350 | 200 | 3550 | 500 | 1500 | 2000 |
|  |  | Torsemide | 15 | 400 | 415 | 375 | 63 | 438 |
|  |  | Total | 3365 | 600 | 3965 | 875 | 1563 | 2438 |
|  | No CKD | Furosemide | 150 | 800 | 950 | 500 | 1500 | 2000 |
|  |  | Torsemide | 35 | 100 | 135 | 375 | 62 | 437 |
|  |  | Total | 185 | 900 | 1085 | 875 | 1562 | 2437 |
| Does not have T2DM | Has CKD | Furosemide | 3350 | 200 | 3550 | 500 | 1500 | 2000 |
|  |  | Torsemide | 15 | 400 | 415 | 375 | 63 | 438 |
|  |  | Total | 3365 | 600 | 3965 | 875 | 1563 | 2438 |
|  | No CKD | Furosemide | 150 | 800 | 950 | 500 | 1500 | 2000 |
|  |  | Torsemide | 35 | 100 | 135 | 375 | 62 | 437 |
|  |  | Total | 185 | 900 | 1085 | 875 | 1562 | 2437 |
|  |  | Total for entire table | | | Furosemide | | | 17000 |
|  |  |  | | | Torsemide | | | 2850 |
|  |  |  | | | Total | | | 19850 |

*Total represents total across the entire population

**Table 4: Absolute risk under each rule in 3-modifier setting**

| Rule | Absolute 1-year risk of mortality |
|---|---|
| Everyone receives Furosemide | 0.52 |
| Everyone receives Torsemide | 0.46 |
| Tailored rule | 0.15 |

This level of stratification exposes a more nuanced effect heterogeneity: individuals with reduced ejection fraction only have lower mortality risk when prescribed torsemide if they also have CKD, irrespective of their T2DM history (Table 3). Here, the tailored rule further reduces the absolute risk from 0.52 to 0.15, which is greater than either of the one-treatment-fits-all rules (Table 4). The absolute risks are slightly different between the one-modifier stratification (0.17) and the three-modifier stratification (0.15), because of the additional modifiers and their impact on the weighting of each stratified risk.

In a clinical epidemiological study, we may have many more modifiers for which average treatment effects are rarely stratified. Higher orders of interactions are less likely to be evaluated without machine learning. In the real-world example below, we investigate the benefits of a tailored rule using machine learning in a higher-dimension setting.

**Precision Medicine Methods**
In higher-dimensional settings, applying machine learning can improve the statistical and computational efficiency of identifying a tailored treatment rule. The key difference between the toy example above and the real-world application below is the use of precision medicine methods to identify the tailored rule. In figure 1, we differentiate the average treatment effect from the precision medicine approach. For the average treatment effect, the recommended rule is a universal rule for all patients. For precision medicine, we generate a tailored rule in a few steps. First, all variables are reduced to those that are most informative to the analysis. Second, the tailored rule is generated by dividing the data into training and test data. The tailored rule is generated on the training data, and evaluated on the test data. Finally, the tailored rule is applied to all the patients to identify the optimal treatment. In our study, each algorithm (efficient augmentation and relaxation learner, random forest, and residual weighted learner) identifies the tailored rule using a conceptual foundation from reinforcement learning.

**Figure 1: Comparison of two intervention rules using a precision medicine approach**

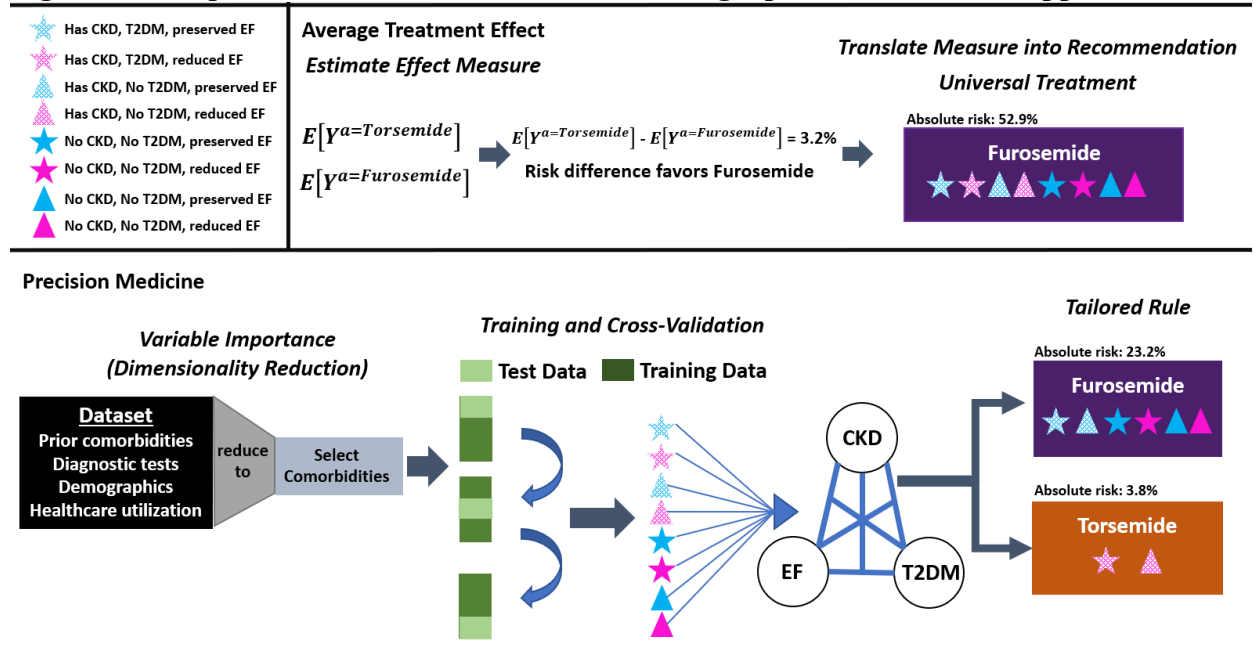

Abbreviations: CKD – chronic kidney disease, EF - ejection fraction, T2DM – type 2 diabetes

*Treatment Rule Estimation*

In reinforcement learning, value functions are key components of the optimization process. Optimal treatment rules are mathematically represented as a map from the patient's baseline history to the treatment decision. Here, treatment A in the binomial setting is ∈ {-1,1}. Leveraging the framework of potential outcomes, the optimal treatment rule is d*(x) where d*(x)= argmax$_{a \in \{-1,1\}}$V(d), where V(d) represents the chosen optimized value function. The value function incorporates the reward, which in this setting is time to one of the two outcomes: all-cause mortality or the composite outcome of all-cause mortality and heart failure readmission. Each precision medicine method used here optimizes a single value function.[6,18,19] The value function in this study was:

$$V(d) = E\left[\frac{Y}{\pi(A;X)} I\{A = d(X)\}\right]$$

In this function, Y represents the reward, A represents the treatment, and X represents the selected characteristics for the rule. This value function is an extension of inverse probability weighting, multiplied by the indicator function capturing if the learner's optimal treatment for that individual matches the treatment received. Estimating the propensity score, π, is a critical step in estimating the value function. In this study, we used random forests to estimate the propensity score for the value function of each learner.

*Random Forests*

Random forests are an established, frequently-used, tree-based ensemble learning method. They produce robust results for classification and regression problems.[19] Random forest algorithms apply bootstrap aggregating (bagging), by selecting a random sample with replacement of the training set and then training a classification tree from the responses in the training set for each of the user-specified samples. At each stage of the learning process, a random subset of the covariates, where a third of the number of covariates are used in each stage. While random forest is a consistently-robust machine learning approach, in precision medicine random forest applications have been outperformed by residual weighted learning.

*Residual Weighted Learning*
Residual weighted learning is one of the latest in a series of weighted learners.[20] The first step of residual weighted learning is the estimation of the conditional expected outcomes given clinical covariates with the most appropriately identified regression model. Second, residual weighted learning calculates the estimated residuals for each outcome by comparing the observed and expected outcomes. If the observed outcome is better than the expected, the residual is positive; otherwise, it is negative. Finally, residual weighted learning identifies the decision function using weights calculated from the first and second step.

*Efficient Augmentation Relaxed Learning*
Efficient augmentation relaxed learning was developed to integrate doubly-robust statistical properties.[18] The key difference between efficient augmentation relaxed learning and other learners is this doubly-robust property. The value function for the former method relaxes model specification assumptions such that you only need to specify either the treatment or the outcome model. An issue that developers of efficient augmentation relaxed learning experienced was that the maximization of the augmented inverse probability weighted estimator value function is not continuous, so they modified the estimator to generate a more appropriate value function (see Zhao et. al for details on the maximization of a concave relaxation).[18]

**Real-world application**
We applied each of these three precision medicine algorithms to identify an optimal loop diuretic prescribing rule for heart failure Medicare beneficiaries. We estimated the optimal treatment rule at discharge following the patient's incident heart failure hospitalization.

*Data sources*
We used a 20% random sample of the fee-for-service Medicare population to estimate optimal treatment rules.[21] This database includes Medicare insurance enrollees, who are aged 65 and older, or are aged less than 65 with certain disabilities or end-stage renal disease. This study was approved by the University of North Carolina at Chapel Hill Institutional Review Board.

*Inclusion criteria*
The study sample included beneficiaries with claims data for at least one calendar month between January 1, 2007 and December 31, 2017.[22] We restricted to beneficiaries over 65 with an inpatient admission with heart failure as the primary admission diagnosis. Additionally, we required beneficiaries to have a loop diuretic prescription within two weeks of the heart failure hospitalization discharge date. We will define the index date as the loop diuretic prescription claim date. As part of the causal framework for optimal treatment rule identification, we identified beneficiaries with no other loop diuretic prescription in the six months prior to the index date. We excluded beneficiaries who received more than one loop diuretic prescription on the index date and beneficiaries with a loop diuretic prescription administered through any route other than an oral tablet. We identified 66,741 beneficiaries who met these criteria.

*Covariates*

We selected several variables for the optimal treatment rule identification: patient demographics, clinically-relevant characteristics (alcohol use, smoking status, frailty), hospitalization history, comorbidities (cardiac, renal, and other cardiorenal associated-conditions), and prescription medications. Ejection fraction was coded using a recently validated claims-based algorithm.[23,24] To adjust for disease severity, we adjusted for loop diuretic dose. Because doses are different between each loop diuretic type, we calculated a dose equivalency where 40 mg of furosemide equals 20 mg of torsemide, which equals 1 mg of bumetanide.[25] All baseline covariates were collected based on claims for the six months prior to the first loop diuretic prescription claim.

*Reward*

The primary reward was increased survival before death by any cause (all-cause mortality) within one year following loop diuretic initiation. We also analyzed time to a composite of heart failure hospital readmission and/or all-cause mortality as a secondary reward. We defined heart failure readmission as anyone with a readmission with a primary diagnosis of heart failure during the year after loop initiation. We used the same ICD-9/10 codes for heart failure that were used for initial identification of heart failure hospitalization. We organized the data using SAS for Windows 9.4 (SAS Institute, Cary NC). We analyzed the data using R (version 3.5.3). [26,27]

*Variable Importance*

Dimensionality reduction is a transformation approach to move from a high-dimensional covariate setting to a low-dimensional covariate setting while preserving the meaningful relationships in the high-dimensional setting. This approach is commonly used when implementing machine learning algorithms. Here, we used a tree-based method to measure variable importance. Importance of each variable is calculated using a calculation for a unitless metric called the Gini impurity, which is a measure of the probability that a randomly selected sample will be incorrectly classified according to the distribution of samples. We chose a leftover, out-of-bag approach where one covariate at a time was removed and remaining covariates were randomly shuffled to calculate importance. This step was repeated until each covariate was removed at least once. The errors were averaged across all trees in the forest. Across 10-fold cross-validation, we identified ten variables with consistently high importance scores (Figure 2), and age had the most importance.

**Figure 2: Variable importance plot**

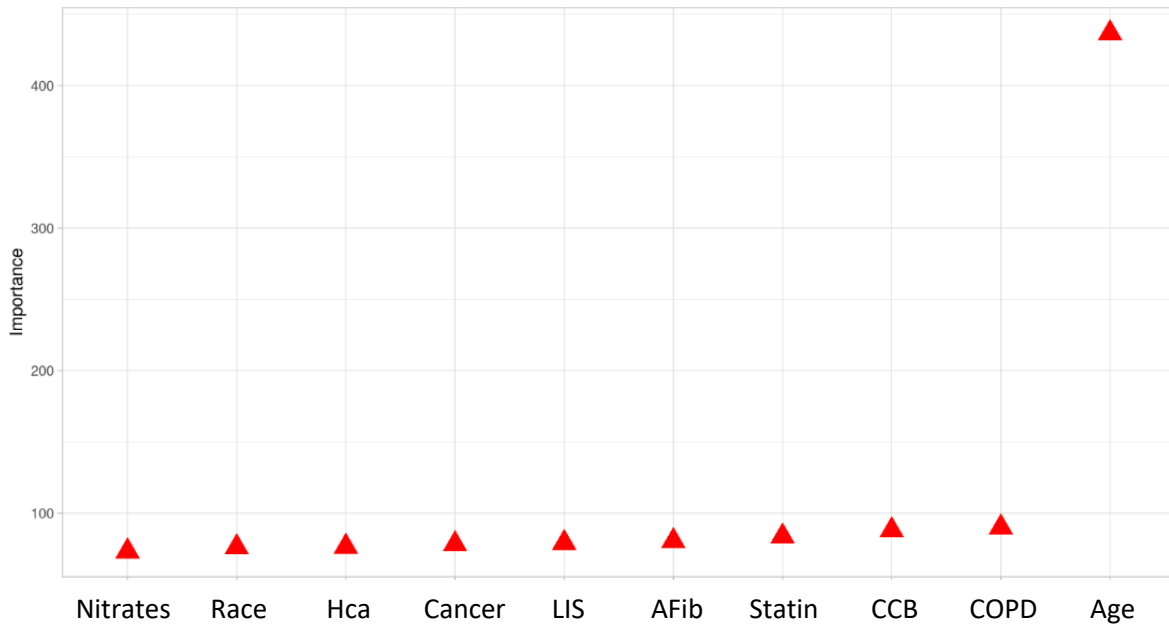

Abbreviations: COPD - chronic obstructive pulmonary disease; Hca - hypercholestoremia; CCB – calcium channel blocker; LIS – low-income subsidy; AFib – atrial fibrillation

*Censoring*

We censored individuals who ended enrollment prior to experiencing an outcome and the end of the study period. To account for censoring, we used recursively imputed survival trees (RIST) to impute the times for these individuals had they not ended enrollment before the end of the study.[28] An extremely randomized tree model was fit to the initial set of training data (half of the data for each drug), where 50 independent trees were fit to the training dataset. The log-rank test statistic was used to determine the best division of the data, and the processing ensured that the final step included at least 1 event. We repeated the imputation 2 times.

*Methods*

We evaluated the expected value function (time to each outcome) for each precision medicine method (Table 3). We used 10-fold cross validation. This means the data were divided into ten groups, one group is held out for the testing the rule set, and the rule is identified in the remaining data (training data). This was repeated so each of the ten groups were used for the test set at least once, and used to train the model nine times. Standard errors were calculated across each of the 10 repetitions. We calculated differences between these expected values and the expected value under a zero-order model, a model where everyone would be prescribed torsemide, to estimate the benefit of applying the precision medicine model vs. a one-treatment-fits-all rule. The best learner is generally identified by the "best" expected value, which, in this study, was the lowest number of days with lowest standard error.

**Results of the real-world application**
Each precision medicine learner performed similarly, as demonstrated by similar expected values and standard errors (Table 5). The greatest treatment optimization was achieved in the composite outcome of heart failure readmission and all-cause mortality within one year. Greatest differences in outcome-free survival were observed under the efficient augmentation relaxed learner.

**Table 5: Expected number of days to each outcome under each treatment rule with corresponding confidence intervals within 1 year, Medicare 2007-2017**

|  | Days alive | Difference | Days alive free of heart failure readmission | Difference |
|---|---|---|---|---|
| **Everyone received Torsemide** | 280.1 | - | 252.6 | - |
| **Residual Weighted Learner** | 289.1 (287.8, 290.5) | 9.4 (3.3, 15.4) | 259.9 (257.5, 262.4) | 7.6 (2.2, 13.0) |
| **Random Forest** | 289.0 (285.1, 292.9) | 7.8 (3.4, 12.1) | 257.4 (253.8, 261.1) | 4.8 (-2.1, 11.7) |
| **Efficient Augmentation Relaxed Learner** | 290.5 (289.7, 291.2) | 10.7 (4.1, 17.3) | 257.0 (255.2, 258.8) | 4.7 (-1.3, 10.6) |

**Discussion**
This study highlights advantages and challenges of precision medicine methods. In the toy example, we demonstrate the value of tailored rules to synthesize complex interactions of effect heterogeneity, particularly when we lack prior knowledge on existing interactions. In the real-world application, we saw that applying a tailored treatment rule results in longer survival times (or heart failure hospitalization-free survival) than one-treatment-fits-all rules.

Each precision medicine method used in this analysis has advantages and disadvantages. This paper was not designed to compare and contrast the statistical advantages of each learner, but rather was focused on introducing these algorithms and their implementation. Efficient augmentation relaxed learner benefits from its doubly-robust feature, allowing for misspecification in one of the two models used to generate the estimator.[18] However, because both models must be specified, efficient augmentation relaxed learner requires more computational effort. Efficient augmentation relaxed learner may be considerably slower in high-dimensional problem settings, to the point that the analysis cannot be feasibly completed if the researchers are limited or do not have access to additional computing power. Additionally, efficient augmentation relaxed learner does not have the benefits (i.e., weighting the outcomes by residuals) of residual weighted learning. Random forest leverages tree-based machine learning to capture complex interaction structures and can answer classification or regression questions.[19] Because random forest involves a large number of decision trees, it is harder to interpret the generated output and outlier tree values. Residual weighted learning generates more stabilized estimates since it uses residuals to weight the errors.[6] However, residual weighted learning can be inefficient in contrast to other two-step methods, when the mean outcome model is correctly specified. An important next step is to feed these methods into an ensemble learner that leverages the strengths of each method, but a superlearner approach still is constrained by the computational needs of the most intensive approach.

These precision medicine methods previously have not been used to identify an optimal treatment for prescribing loop diuretics to manage heart failure. Here, we combined the causal frameworks commonly implemented in epidemiology with a series of precision medicine and machine learning advances not previously combined to generate a rule. When stratified by subgroups, loop-diuretic prescriptions have comparable results using one-treatment-fits-all rule (estimating conditional risks using inverse probability treatment weighted Kaplan-Meier estimator). Despite the absence of stratified subgroup effects, the precision medicine algorithms here still see differences between the tailored and one-treatment-fits-all rules. The reward benefit, additional outcome-free survival, ranged from four to ten days within 1 year. As this is one of the early applications of these methods in pharmacoepidemiology, there is little context to evaluate the importance of the observed outcome-free survival for this study's tailored rules. In other settings, identified tailored rules are sometimes next evaluated in a randomized trial (SMART design).[29] In this example, clinicians may want to consider whether the value of this added time of outcome-free survival warrants a randomized trial. Applying this rule in practice with the aid of decision support tools would demonstrate whether we continue to observe these differences in practice. Additionally, using these methods to generate rules for other treatment decisions in heart failure beneficiaries could provide context for this study's results.

One key application of statistical precision medicine is the development of data-driven decision support tools.[30] Since the rules that are generated from these methods are difficult to translate into human-interpretable text, support tools are the most reasonable way for healthcare professionals to leverage the rules generated by these methods.[4,30] These support tools would translate the actions of these algorithms into discrete steps, while preserving the complexities of the algorithm to the extent possible. While decision support tools for precision medicine are in early stages of development, they will be important platforms to apply these rules to new beneficiaries and to evaluate the performance of those rules. To inform the structure of these tools, we need to examine how these algorithms function relative to our existing decision support procedures. Existing decision support procedures include traditional clinical trials/real-world evidence studies, sometimes compiled into literature reviews or meta-analyses for regulatory or policy agencies.

In addition to the interpretation of results without a decision support tool, the current state of precision medicine methods limits the study results here. First, in each method, we are unable to compare three treatments simultaneously, so we could not include all three loop diuretics of interest. Second, optimal treatment rule identification methods still require further development of appropriate confidence intervals. Reliable confidence intervals cannot be generated using standard approximations of the sampling distributions.[31] Ongoing work is investigating alternative approaches to generate confidence intervals, including leveraging local consistency of the parameters for an optimal treatment rule.[32] While confidence intervals serve as an important summary of random error in most analytic results, the accuracy of previously estimated decision boundaries have been identified and warrant the attention of clinical audiences.[33,34]

The steps outlined in this study can be extended to other epidemiological settings, assuming a causal framework is appropriately fit prior to estimating the optimal treatment rule. While several challenges remain in improving inference, interpretation and decision support tool development, we demonstrate how identifying treatment rules can improve outcomes.

**Appendix**

**A1. Expected number of days to each outcome under each treatment rule with corresponding confidence intervals within 1 month, Medicare 2007-2017**

|  | Days alive | Difference | Days alive free of heart failure readmission | Difference |
|---|---|---|---|---|
| **Everyone received Torsemide** | 29.2 | - | 28.7 | - |
| **Residual Weighted Learner** | 29.3 (29.2, 29.4) | 0.1 (0.0, 0.3) | 28.9 (28.8, 29.0) | 0.2 (0.0, 0.3) |
| **Random Forest** | 29.2 (29.1, 29.3) | 0.0 (-0.1, 0.1) | 28.8 (28.5, 29.0) | 0.0 (-0.2, 0.2) |
| **Efficient Augmentation Relaxed Learner** | 29.3 (29.2, 29.3) | 0.1 (0.0, 0.3) | 28.8 (28.8, 28.9) | 0.1 (-0.1, 0.3) |

**A2. Expected number of days to each outcome under each treatment rule with corresponding confidence intervals within 6 months, Medicare 2007-2017**

|  | Days alive | Difference | Days alive free of heart failure readmission | Difference |
|---|---|---|---|---|
| **Everyone received Torsemide** | 155.3 | - | 145.4 | - |
| **Residual Weighted Learner** | 159.0 (158.3, 159.8) | 3.9 (1.2, 6.5) | 147.8 (146.5, 149.0) | 2.5 (0.2, 4.9) |
| **Random Forest** | 157.3 (155.8, 158.8) | 1.9 (0.1, 3.7) | 146.0 (144.4, 147.6) | 0.6 (-2.0, 3.3) |
| **Efficient Augmentation Relaxed Learner** | 158.5 (158.2, 158.9) | 3.4 (0.8, 6.0) | 146.5 (145.0, 147.9) | 1.2 (-0.8, 3.2) |